\documentclass[multphys,vecphys]{svmult}

\usepackage{makeidx}         
\usepackage{graphicx}        
\usepackage{multicol}        
\usepackage[bottom]{footmisc}

\makeindex             

\newcommand{\rosat}{{\sl ROSAT}}
\newcommand{\chan}{{\sl Chandra}}
\newcommand{\iue}{{\sl IUE}}
\newcommand{\ntt}{{\sl NTT}}
\newcommand{\hst}{{\sl HST}}
\newcommand{\vlt}{{\sl VLT}}
\newcommand{\spitz}{{\sl Spitzer}}
\newcommand{\keck}{{\sl KECK}}
\newcommand{\foc}{{\sl FOC}}
\newcommand{\wfpc}{{\sl WFPC}}
\newcommand{\stis}{{\sl STIS}}
\newcommand{\acs}{{\sl ACS}}
\newcommand{\fos}{{\sl FOS}}
\newcommand{\nic}{{\sl NICMOS}}
\newcommand{\wfpctwo}{{\sl WFPC2}}
\newcommand{\wfc}{{\sl WFC3}}
\newcommand{\hsp}{{\sl HSP}}
\newcommand{\xtins}{{\sl XTINS}}

\begin{document}

  \title*{The HST contribution to neutron star astronomy}


\author{Roberto P. Mignani\inst{1}
}


\institute{
University College London, Mullard Space Science Laboratory (UK) 
\texttt{rm2@mssl.ucl.ac.uk}
}

\maketitle

\begin{abstract}
While  isolated   neutron  stars   (INSs)  are  among   the  brightest
$\gamma$-ray sources, they are among the faintest ones in the optical,
and their study is a  challenging task which require the most powerful
telescopes.  \hst\  has lead  neutron star optical  astronomy yielding
nearly all  the identifications achieved since the  early 1990s. Here,
the major \hst\ contributions in the optical studies of INSs and their
relevance for neutron stars' astronomy are reviewed.

\end{abstract}

\section{Introduction}
\label{sec:1}

Before  the  launch  of  \hst,   optical  studies  of  INSs  were  the
exception. In the first 20 years since the pulsars discovery, only the
Crab and Vela pulsars were identified (Cocke 1969; Lasker 1976), while
optical pulsations  were detected from  an unidentified source  at the
center of SNR  B0540-69 in the LMC (Middleditch  \& Pennypacker 1985),
and  only  a  candidate  counterpart  was  found  for  the  misterious
$\gamma$-ray source  Geminga (Bignami et  al.  1987).  This  score was
expected  to be  considerably improved  by  \hst, thanks  to its  much
larger sensitivity with respect to ground based telescopes, and to the
sharp spatial resolution of the \wfpc\  as well as to the near-UV view
of  the ESA's \foc.   Unfortunately, the  spherical aberration  of the
\hst\ optics affected the execution of most approved proposals, except
for those aimed  at the brightest targets. So, in  the early 1990s the
leadership in  the INSs  optical astronomy was  still in the  hands of
ground-based  observatories, mainly in  those of  the ESO  \ntt\ which
secured the identification of Geminga through the proper motion of its
counterpart, a technique soon become  the standard one, and the likely
identifications  of the  optical pulsar  in  SNR B0540-69  and of  PSR
B0656+14 (see  Mignani et al.   2000).  However, the  refurbishment of
\hst\ in SM-1 (Dec. 2003) brought its performance back to the original
expectations and  gave it a  leading role in INSs'  optical astronomy,
mantained even after  the advent of the 10-m  class telescopes.  Since
\hst\ has provided 8 new  INSs identifications, against the 2 of the
\vlt\  and  the  \keck\  (see  Mignani et  al.   2004),  boosting  the
identification rate by  a factor 4.  This could have been higher
if not for the \foc\ removal in SM-3B (March 2002) and for the
\stis\ failure  (Aug. 2004), which  alone have yielded nearly  all the
\hst\  INSs identifications,  depriving the  telescope of  its near-UV
view.   Thus, \hst\  observations have  opened wide  a  new, important
observing window  on INSs and triggered  the interest of  a larger and
larger fraction of the neutron star community.

\section{\hst\ observations of neutron stars}
\label{sec:2}

Given  their  faintness ($m>25$),  most  INSs  observations have  been
performed in  image mode with the  \foc, the \wfpctwo\  and the \stis,
while \nic\  has been rarely used,  also for being  idle in 1999-2002,
and \acs\ has not been used much before its Jan. 2007 failure.  Due to
their higher near-UV  QE, the \foc\ and the \stis\  have been the best
instruments  for pathfinding identification  programs, while  both the
\wfpctwo\  and  \acs,  with  their  wider field  of  view,  have  been
preferred  for  astrometry (see  Caraveo,  these proceedings).   \fos\
spectroscopy has been performed  only for the brightest objects, while
medium-resolution \stis-MAMA spectra have been obtained in most cases.
Timing observations have  been hampered by the \hsp\ removal in
SM-1,  leaving  \hst\  without  timing  facilities  until  the  \stis\
installation in SM-2 (Feb. 1997), unfortunately idle since Aug.  2004.
Strangely  enough,  \hst\ polarimetry  observations  have been  rarely
performed.

%
%

Most observations  have been  aimed at classical rotation-powered
pulsars.  Multi-band photometry allowed to study their Spectral
Energy  Distribution  (SED) in  detail.   For  the  young Vela  pulsar
(Mignani \& Caraveo 2001) and PSR B0540-69 (Serafimovich et al.  2004)
\wfpctwo\  observations  unveiled   a  power-law  continuum  ($F_{\nu}
\propto   \nu^{-\alpha}$)   confirming,  like   for   the  Crab,   the
magnetospheric origin of the optical emission.  In the near-UV, \stis\
observations  provided the  first spectrum  of the  Crab (Gull  et al.
1998) since the \iue\ one
and of  the Vela  pulsar (Romani et  al.  2005).  Observations  of the
middle-aged  Geminga  and PSR  B0656+14  with  the  \foc\ (Bignami  et
al. 1996;  Pavlov et al.  1997;  Mignani et al.  1998)  and the \stis\
(Kargaltsev et al.  2005; Shibanov  et al.  2005; Kargaltsev \& Pavlov
2007) allowed to identify for the first time a Rayleigh-Jeans spectral
component ($T \sim 10^5$ K),  likely originated by the cooling neutron
star surface.   The detection of near-UV thermal  emission is critical
for neutron star physics.  Coupled with the distance, the joint fit to
the  near-UV--to--X-rays  thermal  spectrum  yields the  neutron  star
thermal   map,  which  is   crucial  to   study  the   neutron  star's
conductivity, hence  its chemical composition  and physical conditions
as well as  the magnetic field topology, and  to constrain the neutron
star  equation of state.   This also  yields a  better measure  of the
surface  temperature which,  coupled with  the neutron  star spin-down
age,  allows  to test  cooling  models.   Last,  studying the  near-UV
thermal emission  is crucial to  explore temperatures too low  for the
X-rays and to constrain neutron star cooling curves above 1 Myr, where
different slopes are predicted, or to pinpoint evidence for re-heating
of the  neutron star core as,  e.g.  for PSR  J0437-4715 whose derived
temperature ($\approx 10^5$ K) largely exceeds any expectation for a 5
Gyr  old INS (Kargaltsev  et al.   2004).  In  the near-IR,  \nic\ has
observed PSR B0656+14 and Geminga (Koptsevich et al.  2001), the first
INSs detected at these wavelengths  after the Crab.  In both cases the
near-IR SED  is consistent  with a power-law  ($\alpha>0$), suggesting
that   the   emission   at   longer  wavelength   is   magnetospheric.
Interestingly,  for PSR  B0656+14 the  power-law steepness  might also
suggest that  the near-IR  emission is  due to a  debris disk  made of
fallback  material from the  supernova explosion  (e.g.  Perna  et al.
2000).   Unfortunately,  \spitz\ observations  could  not resolve  the
pulsar emission from that of  the crowdy background and thus constrain
its  mid-IR  spectrum.  So  far,  polarimetry  observations have  been
performed  only for  the  Crab pulsar.  However, phase-resolved  \hsp\
polarimetry (Graham-Smith et al.  1996) has shown, for the first time,
that  the pulsar polarization  properties are  wavelength independent.
Phase-averaged optical polarization observations of the Crab have been
also performed with the \wfpctwo\  and \acs\ (Mignani et al., in prep)
and are  now in  progress for  PSR B0540-69 and  the Vela  pulsar.  As
shown  by Mignani  et  al.  (2007),  polarization  observations are  a
powerful  diagnostic to  test  neutron star  magnetosphere models,  to
constrain the  pulsar rotation and  magnetic axis, and  to investigate
pulsar/ISM magneto-dynamical interactions. Timing observations have
been performed for nearly all the brightest INSs. \hsp\ discovered the
wavelength dependence of the Crab pulsar light curve profile (Percival
et al.   1993) and provided  a very precise braking  index measurement
for PSR  B0540-69 (Boyd et  al.  1995), the  only one obtained  in the
optical.   The   \stis-MAMAs  detected  for  the   first  time  near-UV
pulsations from Geminga (Kargaltsev  et al.  2005), PSR B0656+14 (Shibanov
et  al.  2005)  and the  Vela pulsar  (Romani et  al.   2005), 
showing the  lightcurve  dependence   on  the   underlying  spectrum.
%

 \hst\ observations have been  fumdamental to understand the nature of
radio-silent INSs whose  high-energy emission is not rotation-powered.
This is the case, e.g. of  the INSs with purely thermal X-ray emission
(\xtins),  discovered  by \rosat\  (e.g.   Haberl  et  al.  2007)  and
originally  thought to  be extinted  radio pulsars,  re-heated  by ISM
accretion.   However, astrometry of  the \xtins\  optical counterparts
discovered  by  \hst\  yielded  space  velocities  too  high  for  ISM
accretion, thus favouring younger ages and a natural cooling scenario.
\hst\  observations have  also allowed  to characterize  their thermal
optical  SED  and  to   build,  thanks  to  the  measured  parallactic
distances, the  surface thermal map,  with a cooler and  larger region
and a warmer and smaller one, emitting the optical and X-ray radiation
respectively.  Together  with the detection of  X-ray pulsations, this
helped  to explain  the claimed,  surprising inconsistencies  with the
neutron  star model cooling  curves which  apparently predict  too low
temperatures for the measured \xtins\ ages.

\hst\ observations have also allowed to resolve for the first time
the structure of  the synchrotron nebulae powered by  the neutron star
relativistic winds  around the Crab  pulsar (Hester et al.   1995) and
PSR B0540-69 (Caraveo  et al. 2001), with morphologies  similar to the
X-ray  ones.   For  the  former,  a  continuous  monitoring  with  the
\wfpctwo\ has  also shown evidence  for an expanding  equatorial wind,
later confirmed in X-rays by  \chan. Very recently, with the \wfpctwo\
we have  found evidence  of variability also  in the  B0540-69 nebula,
which could be attributed to an expanding jet from the pulsar (De Luca
et al. 2007). So far,  evidence for an expanding  pulsar jet was
found  for the  Vela pulsar  only, through  \chan\  X-ray observations
(Pavlov et al. 2001).

%
%


\section{Future perspectives}
\label{sec:3}

Currently,  only the  \wfpctwo\  and \nic\  are  available as  imaging
instruments, both  older than  10 years.  \wfc\  will be  installed in
SM-4, providing a complete near-UV--to--near-IR spectral coverage over
an equally large field of view.  However, while the \wfc\ QE is higher
in the near-IR with respect to \nic, in the near-UV and in the optical
is lower  with respect to  the \stis-MAMAs and \acs.   This encourages
repairing both \stis\  and \acs\ in SM-4.  Furthermore,  \stis\ is the
only instrument suited for INSs  near-UV timing, while \acs\ allows to
carry out polarimetry observations, not possible with the \wfc.  Thus,
an  upgraded and fully-refurbished  \hst\ is  critical to  mantain its
established world-leading role in neutron star astronomy\footnote{
The author thanks C. Nicollier and the SM-1 crew, whose work changed his life}.

%




\printindex
\end{document}